\documentclass[%
 reprint,
 amsmath,amssymb,
pra,
]{revtex4-2}

\usepackage{graphicx}
\usepackage{dcolumn}
\usepackage{bm}

\usepackage{multirow}
\usepackage{graphics}
\usepackage{xcolor}
\usepackage{amssymb}
\usepackage{braket}
\usepackage{longtable}
\usepackage{rotating}
\usepackage{diagbox}

\usepackage{scalerel}
\usepackage[normalem]{ulem}

\begin{document}
\preprint{APS/123-QED}

\title{Dynamics of electromagnetically induced water molecule fragmentation}

\author{Anton V. Bibikov}
\affiliation{Skobeltsyn Institute of Nuclear Physics, Lomonosov Moscow State University, 119991 Moscow, Russia}

\author{Sergei N. Yudin}
\affiliation{Skobeltsyn Institute of Nuclear Physics, Lomonosov Moscow State University, 119991 Moscow, Russia}

\author{Maria M. Popova}
\affiliation{Skobeltsyn Institute of Nuclear Physics, Lomonosov Moscow State University, 119991 Moscow, Russia}

\author{Alexei N. Grum-Grzhimailo}
\affiliation{Skobeltsyn Institute of Nuclear Physics, Lomonosov Moscow State University, 119991 Moscow, Russia}

\author{Elena V. Gryzlova}
\email{gryzlova@gmail.com}
\affiliation{Skobeltsyn Institute of Nuclear Physics, Lomonosov Moscow State University, 119991 Moscow, Russia}

\date{\today}

\begin{abstract}
The development of intense high-energy radiation sources and the improvement of techniques for detecting charged fragments have made possible experiments on multiple ionization of a molecule with registration of the momentum and charge of dissociation products in coincidence. This technique allows to determine (`fix') a molecular geometry at the time of fragmentation and called fixed-in-space molecule.
In this work, the dynamics the  water molecule dissociation fragments resulting from interaction with intense X-ray radiation has been studied. The charge distribution of oxygen ions was calculated, Newton diagrams were constructed for fragments --- protons and the oxygen ion --- for various charge states of the latter, and the released kinetic energy was evaluated. Calculations were performed using the \cite{Artemyev2005} code for parameters close to \cite{Jahnke}. The predictions for the different pulse parameters are done.
\end{abstract}


\maketitle

The molecule of water   is of particular practical interest for the physics of the interaction of small quantum objects with radiation due to its prevalence in the Universe and its significance for the biology. Detailed understanding of the water molecule evolution, i.e. the fractions of created ions and radicals, in an ionizing electromagnetic field is critical for applications such as the study of radiation damage in X-ray diffraction experiments, radiobiology \cite{Nikjoo01011994}, the chemistry of radicals in solutions \cite{Boudaiffa,Garrett} and even for the explanation of some phenomena occurring in the atmospheres of planets \cite{Carlson,Blanc} and comets \cite{Draganic}.

The evolution of an atom or molecule under X-ray pulse begins with the photo\-emission of an electron.
In most cases, the loss of the first electron does not lead to the decay of the molecule, since the remaining electrons create sufficient attractive potential to form a molecular bond, particularly due to the fact that ionization from the inner $K$ and $L$ shells dominates in the high-frequency range. The resulting hole state quickly relaxes due to Auger decay
or fluorescence \cite{Siegbahn1975}. Thus, the first act of ionization triggers a complex chain of competing processes, such as fluorescence, Auger decay, dissociation, and, finally, the Coulomb explosion of the molecule \cite{Lam}, the system may end up completely devoid of electrons \cite{Braube}. The evolution of a sample depends on the field parameters, intensity, duration, polarization, etc \cite{Serkez2018, Gryzlova2023}.


One of the key problems of experiments in the gas phase is the random orientation of the molecule in space, which leads to averaging and, as a consequence, to blurring of many physical  effects. The vector correlation approach, such as coincidence measurements of photoelectrons and charged fragments \cite{Guillemin,Weber,Weber2004}, makes it possible to determine the axis (orientation) of the molecule in space, which, in particular, allows to observe features of the angular distributions of photoelectrons that are impossible in a spherically symmetrical system
\cite{Dorner,Lebech2002,Ullrich_2003}.

 These experiments provide much more detailed information about the physics of the processes taking place, permitting connection electronic and nuclear dynamics \cite{Piancastelli1999,Sankari2020}: for example, it was shown how measuring the angular distributions in a molecular system `fixed' using a coincidence technique allows  to determine the molecular bond length \cite{Fukuzawa}. The use of such methods even permits to determine through which dissociation channel the molecule scattered \cite{Severt,Howard}, paths a way for Auger-Doppler spectroscopy \cite{Travnikova} and
reveals sharp resonance structure
\cite{Bao_2008,Dehmer}

To fix in space a small molecule, the registration  of the ionized fragments may be enough, but to obtain a detailed information on the structural dynamic, one needs to detect as much fragments as possible together with their asymptotic momentum.
This leads us to a Coulomb explosion technique that first was applied for complanar molecules but now it is available even in 3D space \cite{Pitzer2013,Lam,Li2022}. The detection of several fragments in coincidence decreases number of counts and powerful source of electromagnetic radiation, such as XFEL, is needed.

The development of methods for theoretical description of the angular distributions of photoelectrons during the ionization of molecules has started with the work of \cite{Dill1974}, and since then, a variety of methods have been used to describe the states of the continuous spectrum, from R-matrix approach \cite{Moore2011} and the numerical solution of the Schr\"{o}dinger equation \cite{LUCCHESE1986} to the currently very popular XCHEM approach \cite{Marante}.

The spectra of Auger electrons of the water cation H$_2$O$^{+}$ with a $K$-vacancy (SCH, single core hole)  were studied theoretically and experimentally \cite{Siegbahn,Agren1993,Sankari2020}.
Calculations predicted an increase in the cation bond angle to $\approx 120^\circ$ and a lifetime of $\approx 5$ fs. The water dication H$_2$O$^{2+}$ has not been observed in a stable state, but it is efficiently formed during Auger decay of an internal vacancy \cite{Moddeman} or, for example, during double ionization \cite{Scully}. The dissociation of the water dication in lower excited states was studied in \cite{Streeter,Reedy}.
The similarity of the Auger spectra corresponding to the decay of a $1s$ vacancy localized on the oxygen atom for various oxygen-containing molecules was discussed in \cite{Wang2021}.

If the radiation intensity is high, then a dication with a double vacancy on the$ $K-shell (DCH, double core hole) can be formed. The features of the Auger decay of the molecular DCH states were first discussed by \cite{Cederbaum} and have since been the subject of numerous studies, see, for example, the review by \cite{Piancastelli}. The first calculation of the spectra of Auger electrons with a single and double vacancy in the inner shell of water \cite{Inhester} showed that nuclear dynamics in DCH, unlike SCH, has a strong influence on the electron emission spectra, and its lifetime is only about 2 fs.

This work is stimulated by a recent experiment studying the dissociation dynamics of a water molecule, in which the momenta of protons  produced in the Coulomb explosion of the molecule were measured along with the momentum and charge state of the oxygen ion \cite{Jahnke}. In addition to the comparison of our calculations with the available experimental data, we  consider evolution of water fragments and simulate the molecule decay dynamics for pulses with different duration.

Unless otherwise specified, the atomic system of units is used.

\section{Theoretical description}
\label{Methods}

The theoretical description consists of three main parts: (A) calculation of quantum chemical properties of a molecular fragments with a certain charge and configuration; (B) calculations of charge and configuration state of the target in the electromagnetic field as a function of time; (C) simulation of a motion of different molecular fragments. The last part demands both first and second, while these two are developing separately. We consider the probability of any event (photo\-ionization, Auger-decay and fluorescence) independent of a molecular geometry; and chemical properties, such as potential energy surface (PES)  can be calculated for each configuration in advance.

\begin{figure}[h!]
\begin{center}
\includegraphics[width=0.4\textwidth]{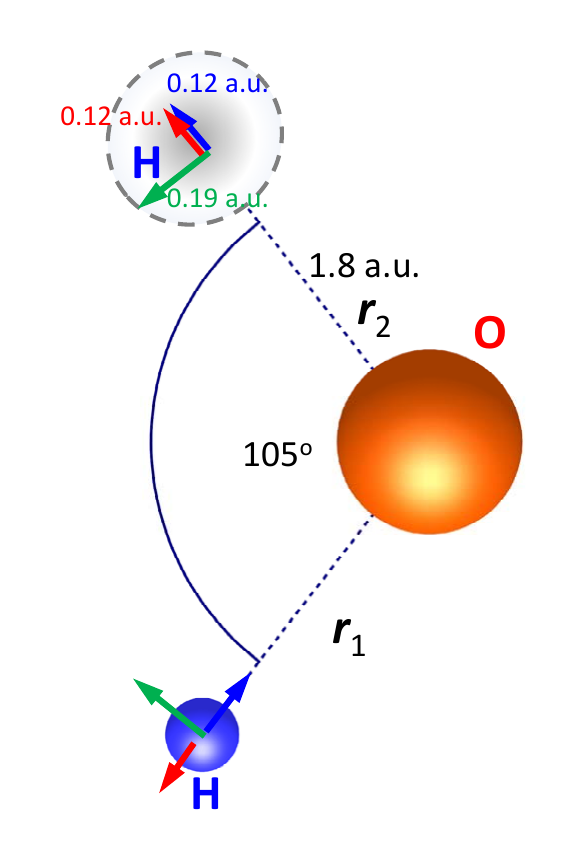}
\caption{The neutral water geometry; shaded area indicates a region  around  the equilibrium available due to the vibrational modes: the green arrows schematically indicate the scissors' mode, the red ones --- symmetrical mode, and the blue arrows --- asymmetrical mode. }
\label{fig:fig0a}
\end{center}
\end{figure}

\subsection{PES of water molecule ions}
\label{ch:PES}

We would like to start with consideration of potential energy surface $\mathcal{P}$ and corresponding vibrational frequencies.  For a planar triatomic molecule, the PES is a function of the two vector's lengths and the angle between them (bond angle). For a water molecule, it is natural to choose the radius vectors from the oxygen atom to the hydrogen atoms ($r_{1}$, $r_{2}$) and angle $\theta$ between them:

\begin{equation}\label{eq:PSPE}
\begin{split}
&
\mathcal{P}  =  V\left(r_{1},r_{2},f\right)\,, \\
&
f=\cos\theta=\frac{({\bf r}_{1}\cdot{\bf r}_{2})}{r_{1}r_{2}}\,.
\end{split}
\end{equation}
Knowing potential energy as a function of the coordinates, one can estimate zero-order vibrational frequencies and un\-harmonic correction to them \cite{Bounouar}.

For the water cation and dication H$_2$O$^{+/2+}$, we performed PES calculations using the original code \cite{Artemyev2005,PhysRevC.81.024610,PhysRevC.86.014608,PhysRevC.88.034608,bibikov2023} tested on calculations of BeO, Be(OH)$_2$ crystals and the molecules Be@C$_{36}$. The energy of electrons in the field of three heavy nuclei of the H$_2$O molecule was calculated in the unrestricted Hartree-Fock (UHF) approximation, taking into account electron-electron correlations in the second order of perturbation theory (MP2). An extended set of molecular basis functions quadruple zeta aug-cc-pVQZ, which takes into account correlations and valence polarization, was used~\cite{Dunning,Schuchardt}. To control the calculation accuracy, some of results were compared with ones obtained with the GAMESS software package using the ROHF method, taking into account correlations by the MP2 method in the aug-cc-pVQZ basis.

\begin{figure*}[tbph]
\begin{center}
\includegraphics[width=0.4\textwidth]{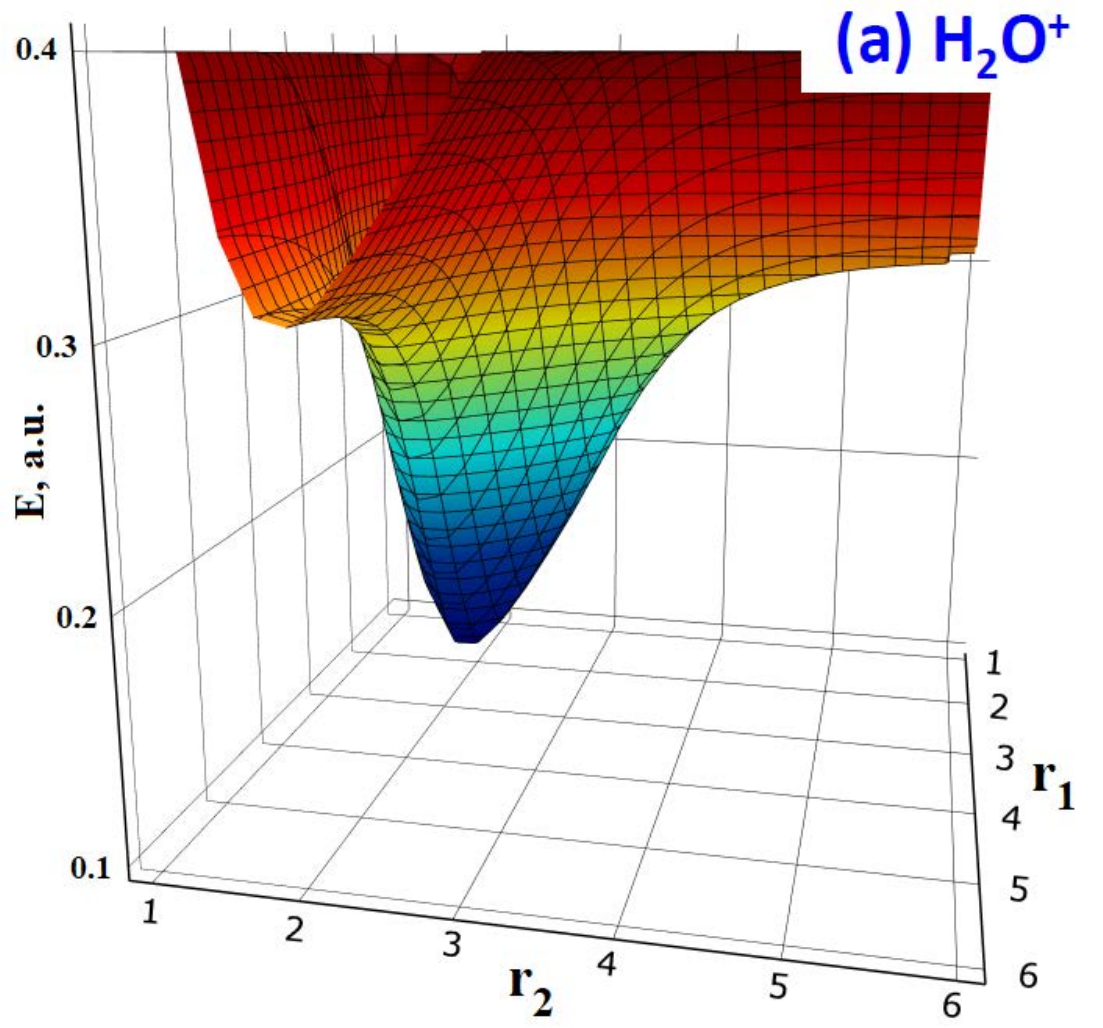}
\includegraphics[width=0.4\textwidth]{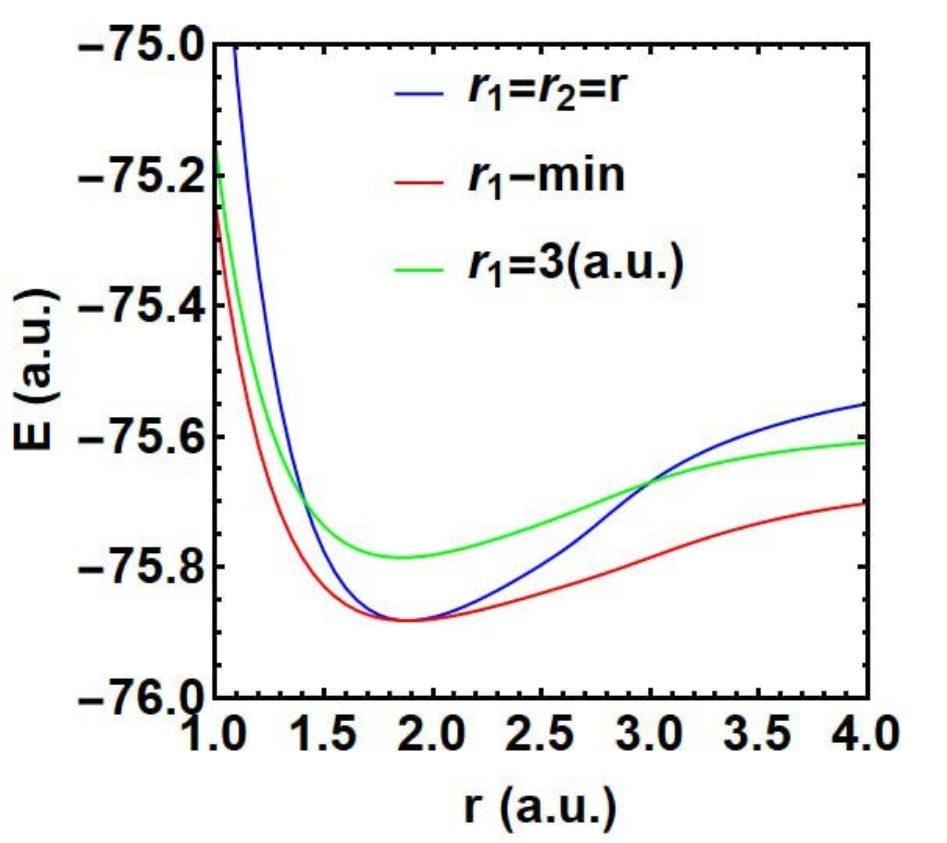}\\
\includegraphics[width=0.4\textwidth]{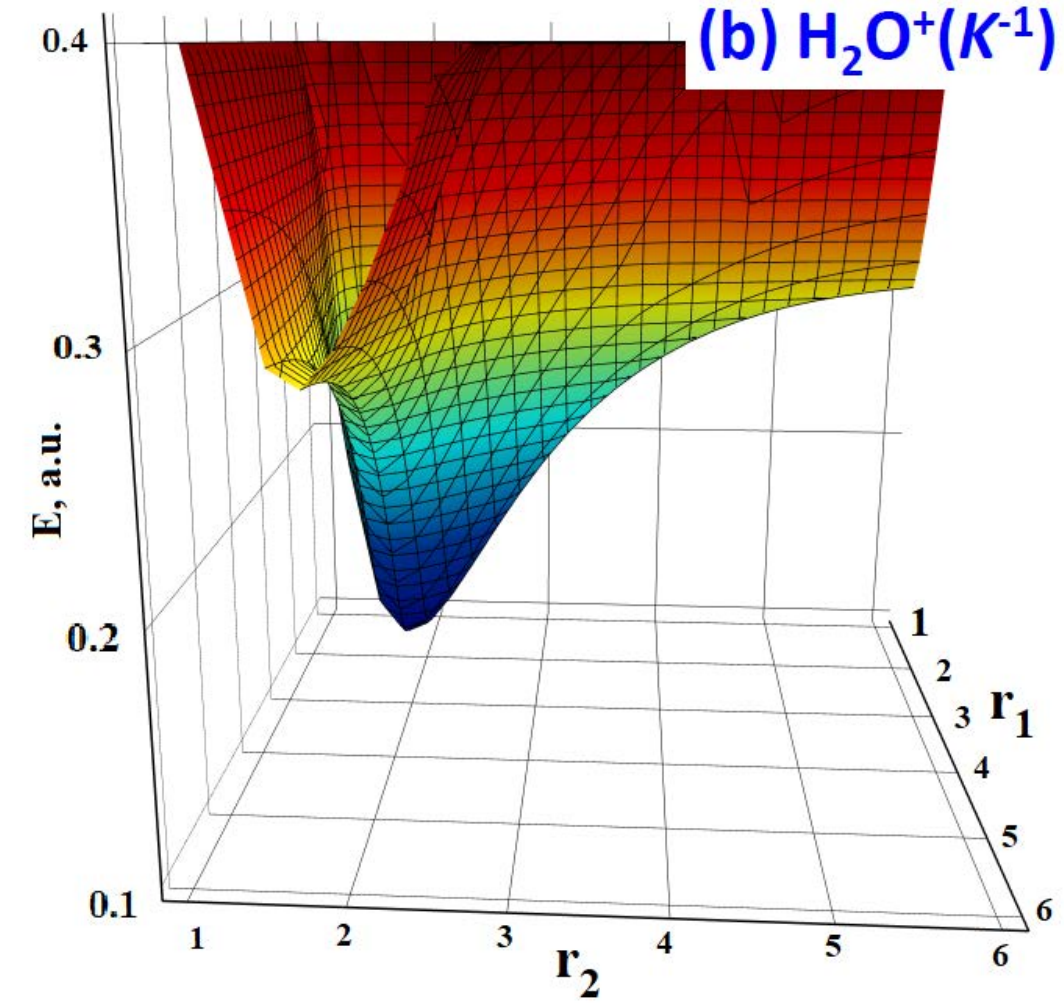}
\includegraphics[width=0.4\textwidth]{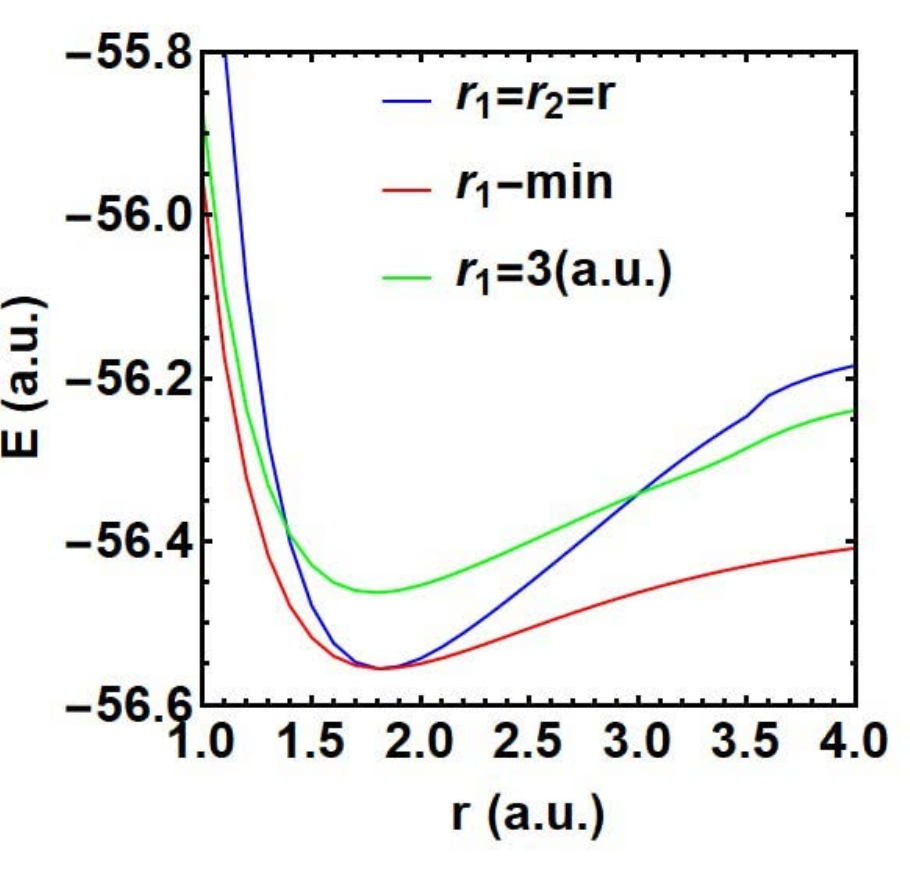}
\caption{PES of water cations as a function of distance between oxygen and hydrogen nuclei ($r_1$ and $r_2$) at the bond angle $\theta=110^{\circ}$. The row (a) presents data for the ground state, the row (b) --- for SCH state. On the right panels, there are sections of the PES with the planes: $r_1=r_2$, one proton is placed at the energy minimum $r_1=r_{min}$, and $r_1=3$~a.u. The 3D plots are shifted for better visualization of the energy minimum.}
\label{fig:fig2a}
\end{center}
\end{figure*}

In Fig.~\ref{fig:fig2a}, there are calculated PES of a water cation in the ground state and with $K$-vacancy localized on oxygen nucleus, i.e. single core-hole state. PES is symmetrical with respect to the line $r_1=r_2$ and manifests a minimum near $r_1=r_2=2$~a.u. The surfaces are plotted for the bond angle $\theta=110^\circ$ which is close to the equilibrium one  of a neutral water  and is slightly less than equilibrium angle of a cation in the ground and SCH-state  (by 4$^\circ$ and 13$^\circ$, correspondingly). The difference of PESs corresponding to the equilibrium bond angle of neutral molecule and cation is less than 0.02~eV. The SCH cation manifests a less pronounced barrier than the cation in the ground state. That means that, despite the fact that the inner shell does not participate in the formation of a chemical bond, a hydrogen nucleus (proton) leaves excited cation easier.

For the neutral water, scissors' mode (1608 cm$^{-1}$) is twice less rigid than symmetrical (3673 cm$^{-1}$) and asymmetrical (3807 cm$^{-1}$) ones. Generally, for water cation all frequencies decrease: 1431~cm$^{-1}$, 3252~cm$^{-1}$, 3314~cm$^{-1}$ for the ground state and  1249~cm$^{-1}$, 3404~cm$^{-1}$ and 3527~cm$^{-1}$ for the SCH state, correspondingly. Our calculations are in reasonable agreement with the experimental data \cite{Sankari2020}, namely $1645$, $3831$, $3944$ cm$^{-1}$ for neutral water, and $1121$, $3710$ cm$^{-1}$ for the SCH cation (the symmetrical mode was not extracted experimentally). A slight difference between the equilibrium bond angle of the neutral molecule and cation is also in agreement with experiment and calculations presented in the literature \cite{Sankari2020}.

\begin{figure*}[tbph]
\begin{center}
\includegraphics[width=0.35\textwidth]{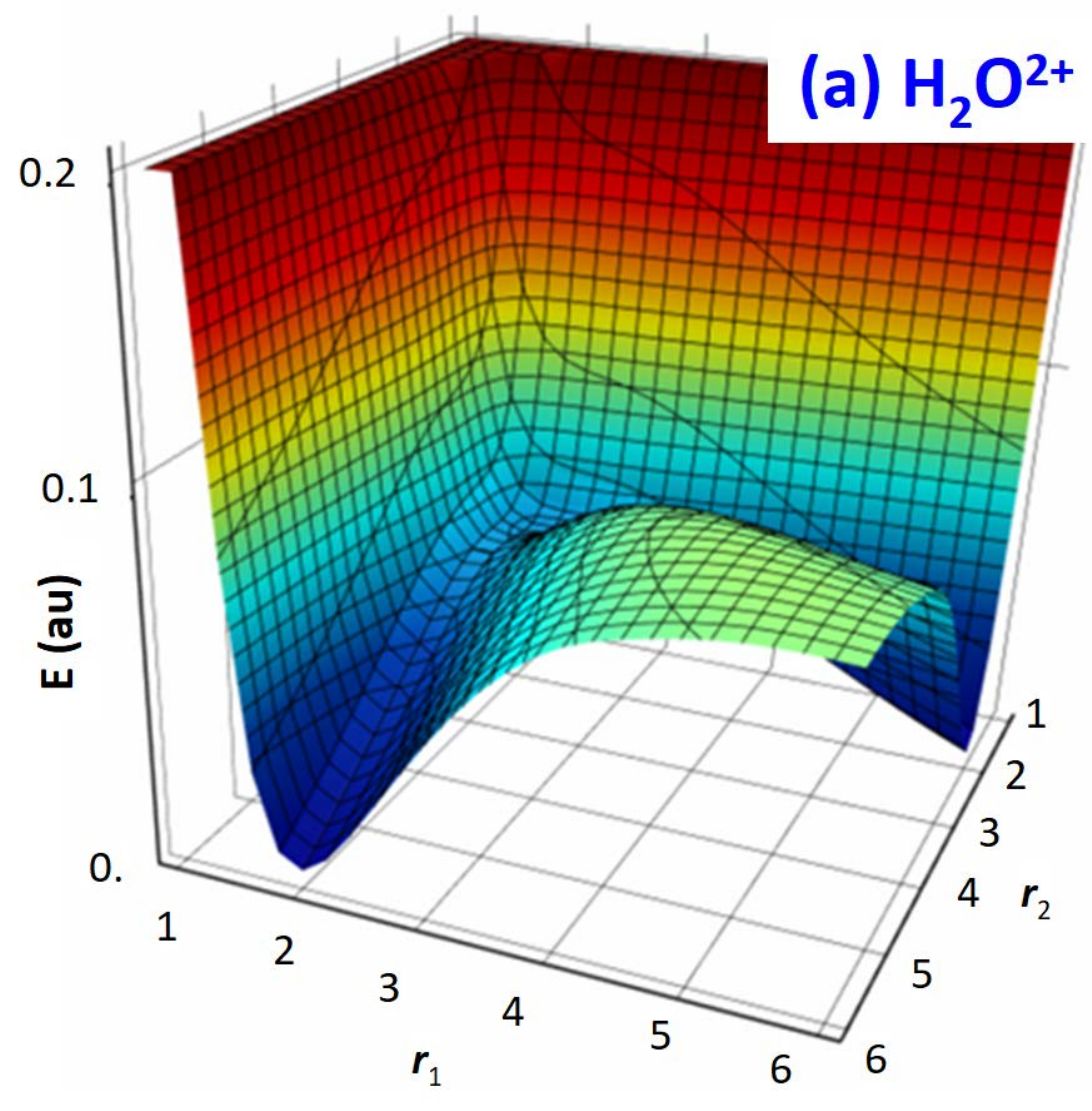}
\includegraphics[width=0.35\textwidth]{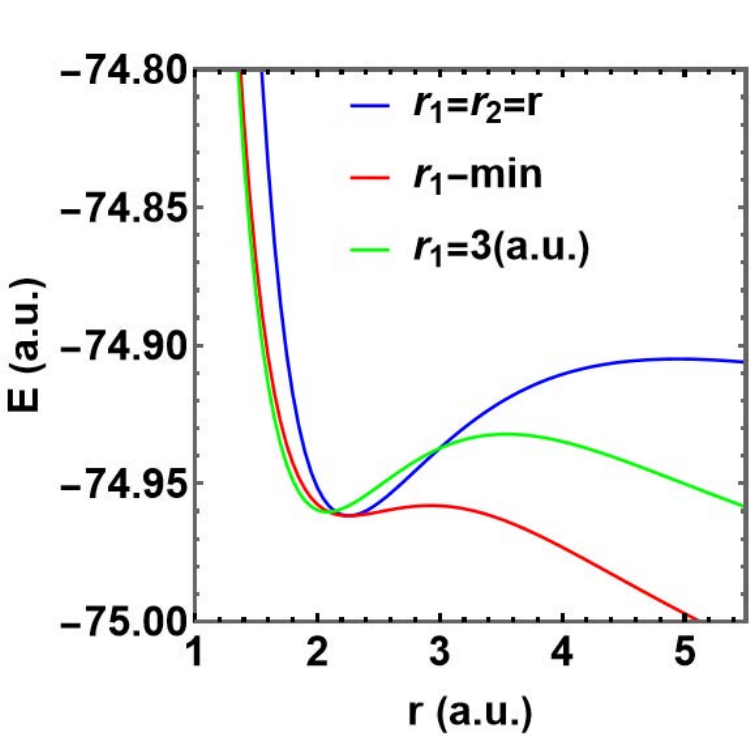}
\includegraphics[width=0.35\textwidth]{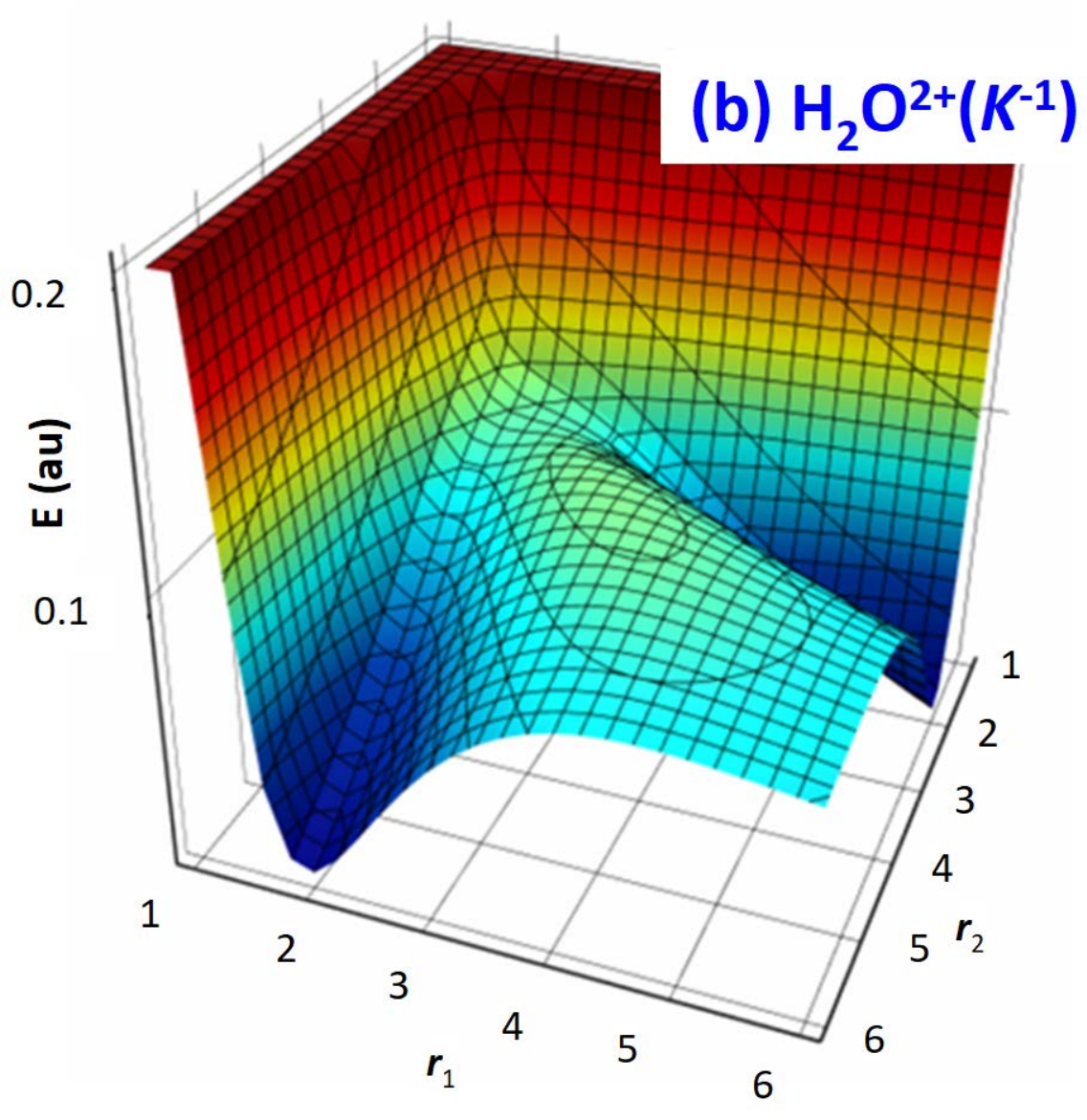}
\includegraphics[width=0.35\textwidth]{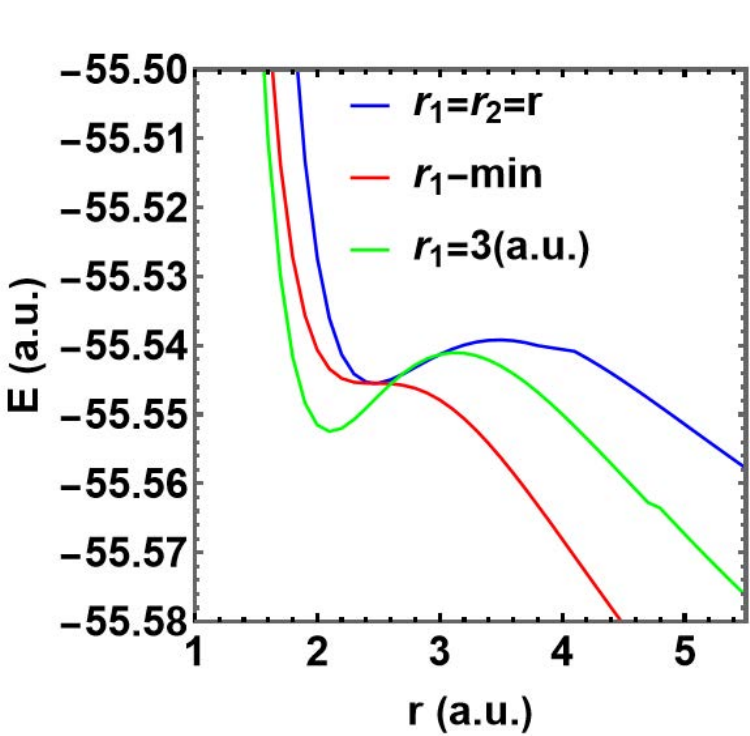}
\includegraphics[width=0.35\textwidth]{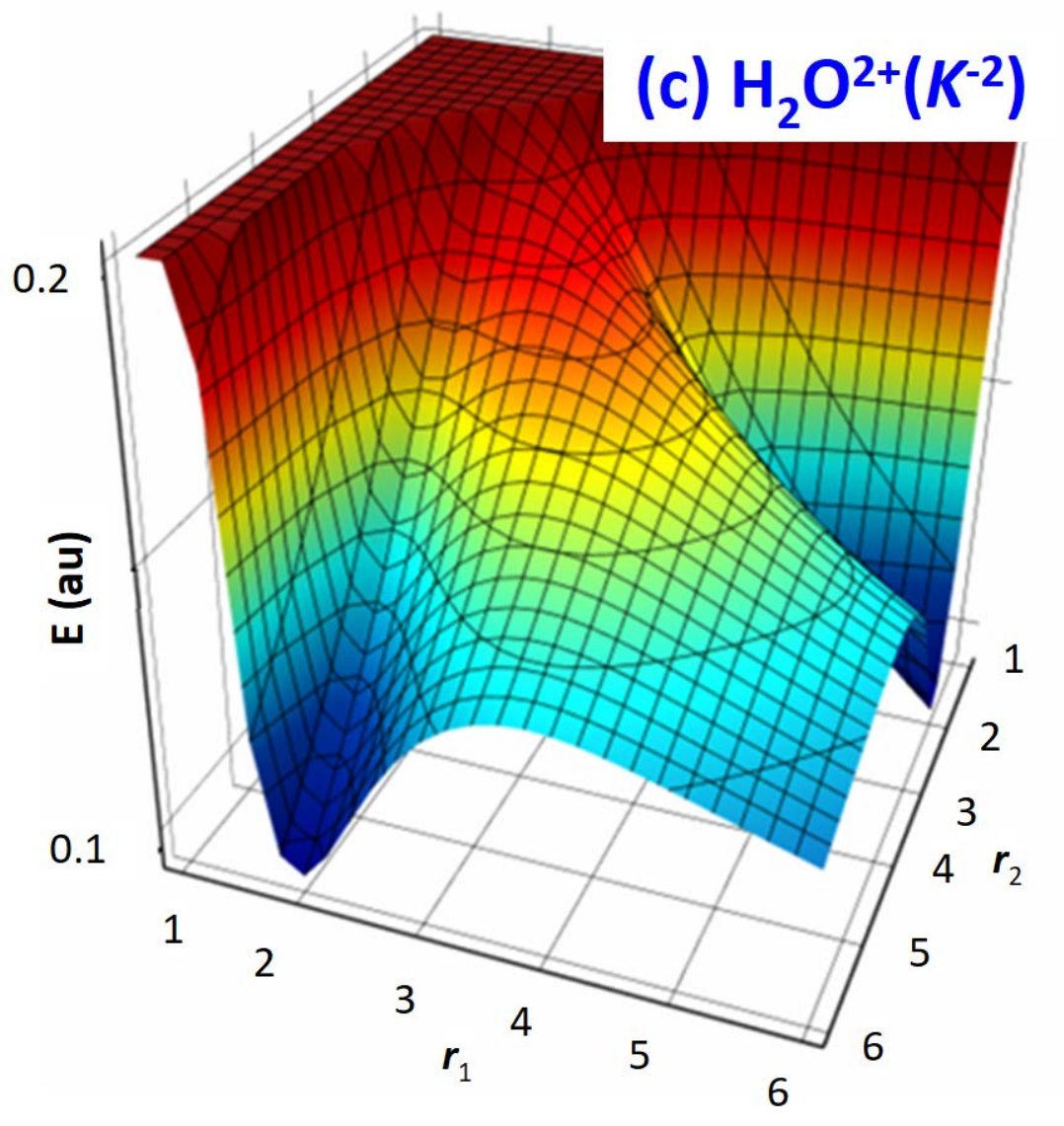}
\includegraphics[width=0.35\textwidth]{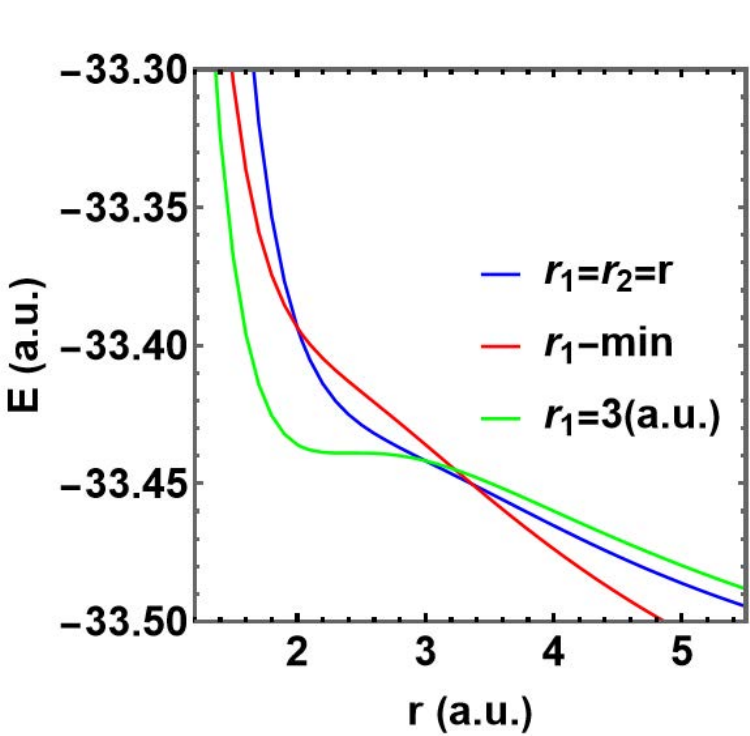}
\caption{The same as in Fig.~\ref{fig:fig2a} but for a water dication. The row (a) presents data for the ground state, the row (b) --- for SCH, the row (c) --- for DCH state.}
\label{fig:fig2b}
\end{center}
\end{figure*}

In Fig.~\ref{fig:fig2b}, there are calculated PESs of a water dication in the ground state, with single $K$-vacancy localized on oxygen nucleus (SCH), and with double $K$-vacancy (DCH state). The ground and SCH states form a minor barrier for the symmetrical dissociation $r_1=r_2$ and even more tiny for asymmetrical dissociation, but there is no  barrier for the asymmetric dissociation along a trajectory corresponding to a more stable position of the remaining proton $r_i=2$~a.u. The DCH state does not form  barriers for any dissociation channels, which is consistent with the conclusions of \cite{Gervais}. The different curves in Fig.~\ref{fig:fig2b} may be tentatively interpreted  as corresponding  to the different predominant dissociation channels:  O+2H$^+$ (the blue curves correspond to symmetrical deformation  at $\theta=180^{\circ}$ in panels a,b), OH$^+$+H$^+$ (the red curves correspond to the most stable position of the remaining proton) and O$^+$+H$_2^+$ (cannot be shown in terms of $r_1\,,r_2$ coordinates). The measurements  \cite{Pedersen2013} manifested that for the ground state of dication,  dissociation into first two channels  is comparable and four times more efficient than into the last channel.

For both SCH and DCH of water dications, the examination of vibrational mode is not relevant for forthcoming discussion  because there is no barrier. For the ground state of dication, different modes are presented in Tab.~\ref{tab:tab1}. The harmonic approximation of the extracted frequencies is in good agreement with those calculated in the GAMESS software, but the cubic and further corrections rearrange the frequencies in a dramatic way: stiffness of symmetrical and asymmetrical modes decreases, while one of scissors mode increases, as a result, asymmetrical mode turns out to be the lowest.
\begin{table}
    \centering
    \begin{tabular}{c|cc|cccc}
   Mode& \multicolumn{2}{c|}{GAMESS} &  &   &  &   \\
       &   harm & anh & harm & anh & cub & qu  \\
      scis & 671  &  662  & 677 & 729 & 13 & 38  \\
sym  &  1663  &  1355 &  1679  &  957 & -998 & 276 \\
asym & 1129  &  869  & 1155 & 533 & -1111 & 489
    \end{tabular}
    \caption {The frequency of different modes of the  water dication in ground state.}
    \label{tab:tab1}
\end{table}

The scissoring vibrations in this context can be interpreted as a relaxation of the bond angle from the angle of neutral water or its cation to the equilibrium bond angle for the water dication (which is $180^\circ$). For the considered parameters of electro\-magnetic pulse, the bond angle, on average, makes no more than one half of oscillation before fragmentation.

\subsection{Charge and configuration evolution}
\label{sec:oxigen}

Under strong XUV field, oxygen  can form ions from O$^{+}$ to O$^{8+}$, and exotic “hollow” configurations can be formed with a higher probability than the ground states of a given ion \cite{Serkez2018}.
It should be emphasized that with increasing ion charge, generally, probabilities of Auger decay, fluorescence and photo\-ionization tend to increase.

The charge configuration based on the probability of transition from the current state to the next is determined at each time interval  by the Monte Carlo method: by generating a random number with a constant probability $P(a\!\rightarrow \!b)=w_{a\rightarrow b}dt$ for Auger decay and fluorescence, and with the time-dependent photo\-ionization probability $P(a\!\rightarrow \!b)=j(t)\sigma_{a\rightarrow b}dt$. Here $w_{a\rightarrow b}$ is the transition rate from configuration $a$ to configuration $b$, $\sigma_{a\rightarrow b}$ is the photo\-ionization cross section, $j(t)$ --- flux density of incident radiation. This way, the number of the trajectories required for collecting statistics (about 100,000\,--\,450,000) is accumulated.

To evaluate the probability of a particular transition (photo\-ionization, Auger transitions, radiative transitions), we  apply an analogue of a genealogical scheme using the probabilities of atomic transitions \cite{Coville1991}.
The simplified approach to transitions rate calculation is considered justified when one studies ionization by high-frequency radiation, where the probabilities of the processes are not affected by any continuum structures  \cite{Cederbaum,Lunin2015,Moribayashi2008}.

The probabilities of ionization of various shells of hydrogen, oxygen and its ions were obtained by the Herman-Skillman algorithm \cite{Herman} and are in full agreement with the calculations \cite{Yeh1951,Son2011}. Auger decay  with participation of two 2$s$ electrons is denoted as ($LL$), involving both 2$s$ and valence electrons --- as (LV) and  involving two valence electrons --- as ($VV$) (as well as fluorescence probabilities) were taken from \cite{Moribayashi2008}. The Auger decay rates of the SCH state used in this study ($7.01\cdot10^{-4}$ ($LL$), $1.36\cdot10^{-3}$ ($LV$), $2.04\cdot10^{-3}$ ( $VV$)) and DCH states ($2.16\cdot10^{-3}$ ($LL$), $4.62\cdot10^{-3}$ ($LV$), $6.39\cdot10^{-3}$ ($VV$)) differ from those given in \cite{Inhester}: $1.51\cdot10^{-4}$, $1.4\cdot10^{-3}$, $3.8\cdot10^{-3}$, and $7.65\cdot10^{ -4}$, $3.6\cdot10^{-3}$, $11.6\cdot10^{-3}$, however, the total width for these two calculations is in reasonable agreement. The systematically smaller width of the ($VV$) channel is associated with the neglect of interatomic Auger decay in our model. We performed the calculations both with presented and  with molecular Auger rates from \cite{Inhester}, and turned pit that except KER (Fog.~\ref{fig:fig4}) all other results are insensitive to this difference.

In the present study, we use pulse with the envelope defined by a single Gaussian or a sum of Gaussian functions:
\begin{eqnarray} \label{eq:pulse}
j(t) & = &  j_0 G(t,\sigma) = \frac{j_0}{\sqrt{2\pi}\sigma} \exp(-t^2/2 \sigma^2) \,,
\end{eqnarray}
where $j_0$ is the amplitude of the flux density,  and full width half-maximum FWHM$=2\sqrt{2\ln 2}\sigma$  defines pulse duration. The characteristic duration of pulses generated by free electron lasers are about 50\,--\,100~fs \cite{Allaria2012,Finetti2017}.
The standard characteristic of an X-ray pulse is the integral photon flux in the pulse (fluence $F$). For pulse (\ref{eq:pulse}) the overall fluence is:
\begin{eqnarray}
   F&=&\sqrt{2\pi}\,j_0\sigma/\omega\,.
\end{eqnarray} \label{eq:fluence}

\subsection{Dynamics of the charged fragments dissociation}
\label{wec:DQF}

The first ionization initiates a chain of events that most likely leads to a subsequent increase in charge. To access the dynamic of the charged target, we solve the  equations of classical motion in a potential:
\begin{eqnarray} \label{eq:wf_decomp10}
\frac{d\bm p_i}{dt}&=&-\partial_{r_i}\mathcal{P}\,, (i=O,1,2)\,;
\end{eqnarray}
This common equation in the chosen coordinates frame ($r_1\,,r_2\,, f\equiv\cos\theta$) takes the form:
\begin{eqnarray} \label{eq:wf_decomp1}
\frac{d\bm p_{1}}{dt}&=&-\partial_{r_1}\mathcal{P}\frac{\bm{r}_{1}}{r_{1}}-\partial_{f}\mathcal{P}\left(\frac{\bm{r}_{2}}{r_{1}r_{2}}-\frac{(\bm{r}_{1}\bm{r}_{2})}{r_1r_2}\frac{\bm{r}_{1}}{r_1^2}\right)\,;\\
\frac{d\bm p_{2}}{dt}&=&-\partial_{r_2}\mathcal{P}\frac{\bm{r}_{1}}{r_{2}}-\partial_{f}\mathcal{P}\left(\frac{\bm{r}_{1}}{r_{1}r_{2}}-\frac{(\bm{r}_{1}\bm{r}_{2})}{r_1r_2}\frac{\bm{r}_{2}}{r_2^2}\right)\,;\\
\frac{d\bm p_0}{dt}&=&-\frac{d\bm p_{1}}{dt}-\frac{d\bm p_{2}}{dt}\,.\label{eq:wf_decomp4}
\end{eqnarray}
Here for cation and dication $\mathcal{P}$  is the PES (\ref{eq:PSPE}) defined according the previous paragraph, while for higher charged targets $\mathcal{P}$ reduces to the Coulomb repulsion potential. As a photo\-electron leaves the system very quickly, we suppose the switching between PESs happens instantaneously at the time-step corresponding to a change of either charge or configuration  generated by Monte-Carlo method.

When electron leaves the system either as a result of ionization or decay, the system gets an additional  randomly oriented momentum $\sim ~6$ a.u. The momentum transferred by the knocked-out electron to the oxygen atom cannot cause oscillations in the CMF (center-of-momentum frame) because corresponding energy $2\cdot10^{-4}$ a.u. is more than orders of magnitude lower than the energy of the first excitations of water associated with the motion of nuclei.

As neutral water molecules before the first ionization event experience zero-point oscillations relative to the equilibrium position (see Fig,~\ref{fig:fig0a}), the initial conditions for the numerical solution of system (\ref{eq:wf_decomp1}\,--\,\ref{eq:wf_decomp4}) are the randomly generated coordinates and momenta of the nuclei
 in H$_2$O corresponding to zero-oscillations for each mode as harmonic oscillator (see part A. PES of water molecule ions).

Ionization into a triply-charged state leads to direct Coulomb scattering of charged fragments of the molecule. 
At this stage of the numerical solution, each fragment is given a charge of +1, thus, the  channel of dissociation into O$^{2+}$ + H$_2^+$ turns into O$^+$ and 2H$^+$ automatically. Generally the first channel may be distinguish by means of examination the the relative distance $\bm r_1-\bm r_2$ (small enough) but this situation is strongly prohibited by the PES and contributes very few percents of events. A possibility of further charge and configuration evolution of an oxygen ion remains  till 150\,--\,200 fs after the end of the electromagnetic pulse.

\section{Results and discussion}
\subsection{General overview of the observables}

In this paragraph, we consider experimentally available data  and compare our calculations with \cite{Jahnke}.
The solution  of Eqs.~(\ref{eq:wf_decomp1}\,--\,\ref{eq:wf_decomp4}) provides us with fragment's momenta as a function of time, particularly at the asymptotic limit whem a fragment reaches a virtual detector, and aggregated value of the total number of events with a given total kinetic energy of fragments --- the kinetic energy release (KER).  More precisely, the quantity being measured is the number of events as a function of KER. The another measured quantity is charge distribution of oxygen ions.

\begin{figure}[h!]
\begin{center}
\includegraphics[width=0.45\textwidth]{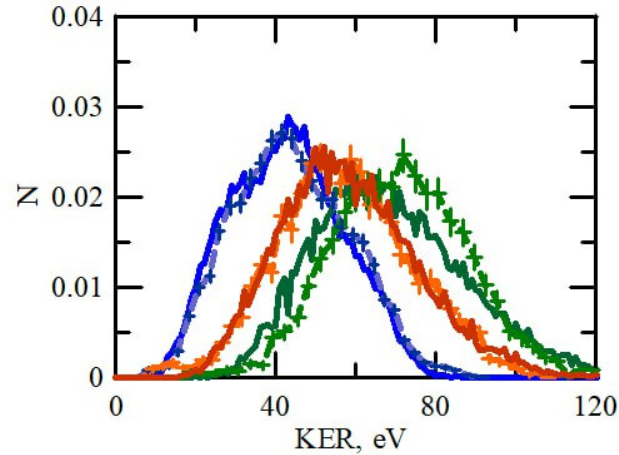}
\caption{The KER measured experimentally \cite{Isinger2019} (crosses) and calculated for the single-Gaussian (\ref{eq:pulse}) pulse (dashed line) with FWHM=20 fs and for double-Gaussian (\ref{eq:pulse2}) pulse (solid line, see parameters in the text). Color indicates the  charge of the oxygen ion.}
\label{fig:fig4}
\end{center}
\end{figure}

 The measured parameters are quite sensitive to the form of the pulse. Apparently: if intensity increases rapidly, the target gains charge quickly and obtains more kinetic energy due to a stronger electrostatic repulsion. In the opposite case, if fragments get additional charge when they are well separated, the energy is lower. Different pulse forms and parameters were considered in order to obtain KER closest to the measured in \cite{Jahnke} (see Fig.~\ref{fig:fig4}). The best results were obtained for
\begin{eqnarray} \label{eq:pulse2}
j(t)=j_0 [ c_1 G(t,\sigma_1) + c_2 G(t,\sigma_2) ]\,,
\end{eqnarray}
with FWHM$_1$=10~fs and FWHM$_2$=40~fs and intensity (or fluence) ratio 0.4 and 0.6, i.e. $c_1/c_2=2/3$. Ions of lower charge correspond to curves shifted to the low-energy region, and ions of higher charge correspond to curves shifted to the high-energy region. The difference in the acquired kinetic energy for ions with $\Delta Z=2$ is of the order of 10~eV.

KER is an integral parameter averaging any angular dependency. To look in deeper details, one has to consider angle-energy dependencies. Here we consider a Newton diagram and an angle correlation diagram.
The Newton diagram presents the final momentum of the protons in a coordinate system in which the oxygen ion has momentum along the $n_x$ axis.
The angle correlation diagram depends on angles between each proton and oxygen.

\begin{figure}[h!]
\includegraphics[width=0.45\textwidth]{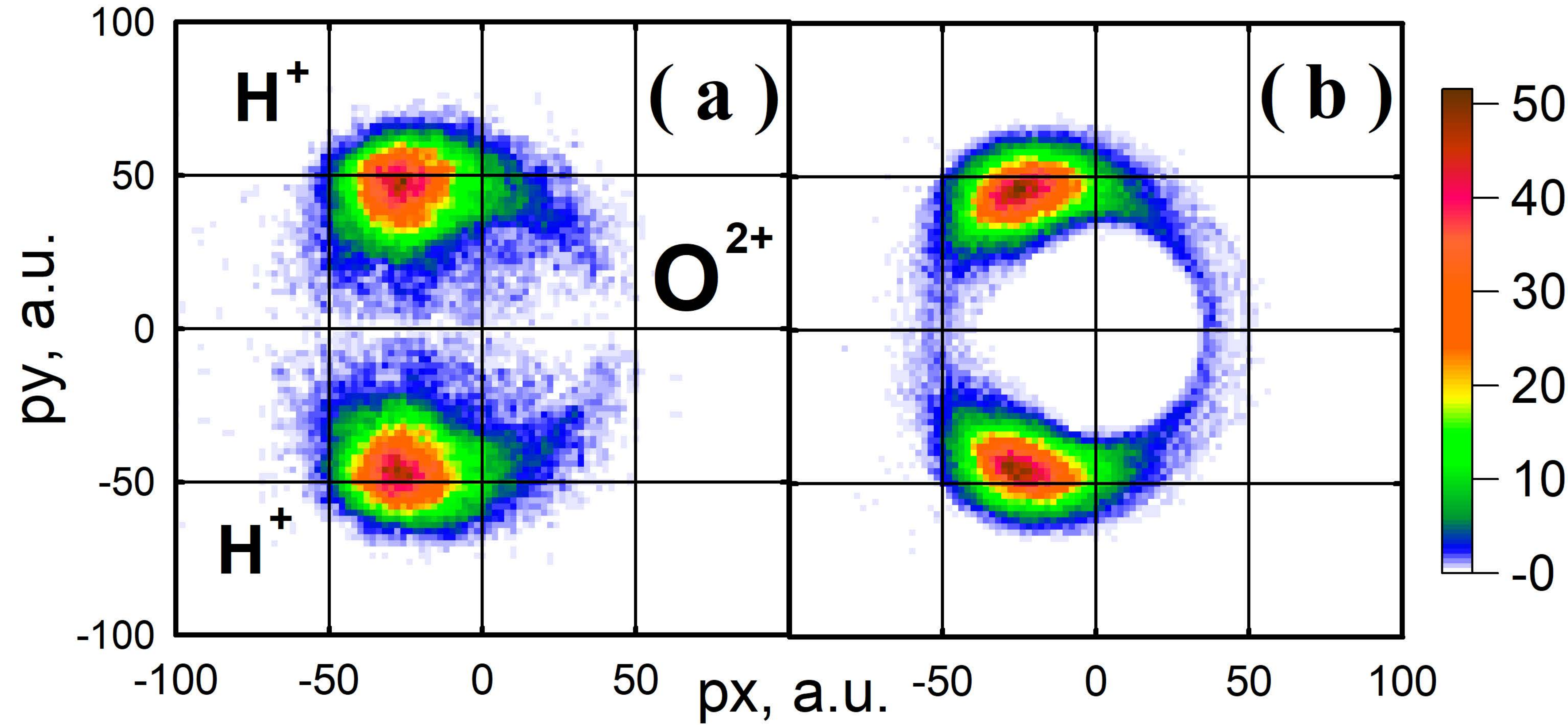}
\includegraphics[width=0.45\textwidth]{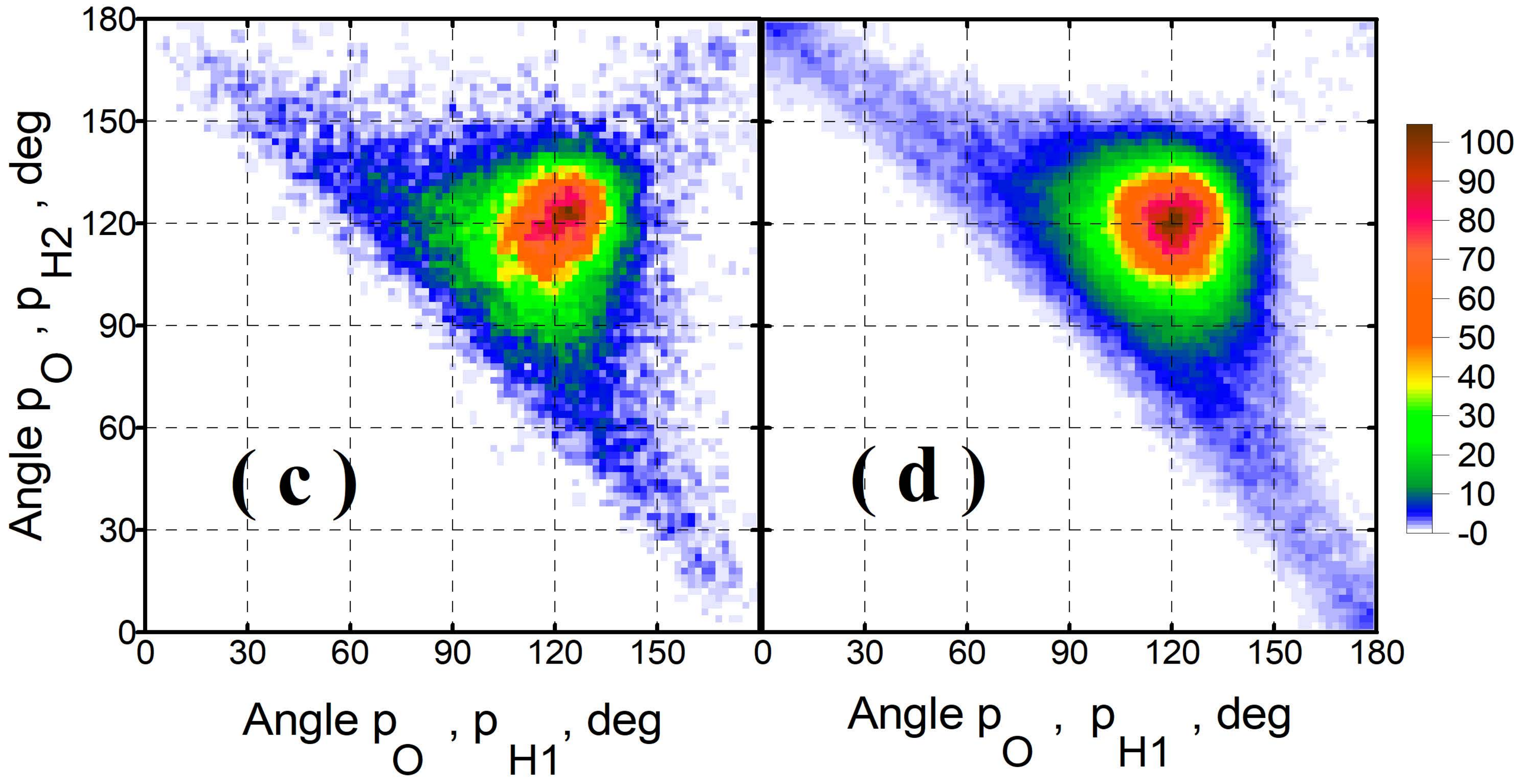}
\begin{center}
\caption{The Newton diagram (a,b) and the angular correlation between protons (c,d) calculated (b,d) and measured  in \cite{Jahnke} (a,c). The calculations were performed for the pulse (\ref{eq:pulse2}).}
\label{fig:Newton}
\end{center}
\end{figure}

In Fig.~\ref{fig:Newton}, we show the Newton diagram for pulse (\ref{eq:pulse2}) for the protons detected in the coincidence with doubly charged oxygen ion O$^{2+}$ and corresponding angular correlation function, accompanied with the experimental data. The overall agreement between theoretical and experimental data is good. Both presentations of the data manifest a bright maximum resulted from  the three-body Coulomb explosion of the molecule ($p_1\approx p_2\,, \theta\approx 110^\circ$) and intricate tails corresponding to the fragmentation from the
unrolled and asymmetric geometries. The bright   blue inner arc in the right side is formed due to the dissociation with a minimal possible energy. It is formed by the first dissociated proton, while the second proton possesses a higher and more diffused energy. The diagonal line $\{180^{\circ}, 0^{\circ}\}-\{0^{\circ},180^{\circ}\}$ in the panels (c,d) corresponds to protons emitted in the opposite directions with practically motionless oxygen ion.  Below we consider the dynamic of process in deeper details.

\subsection{Charge and configuration trajectories}

\begin{figure}[h!]
\begin{center}
\includegraphics[width=0.4\textwidth]{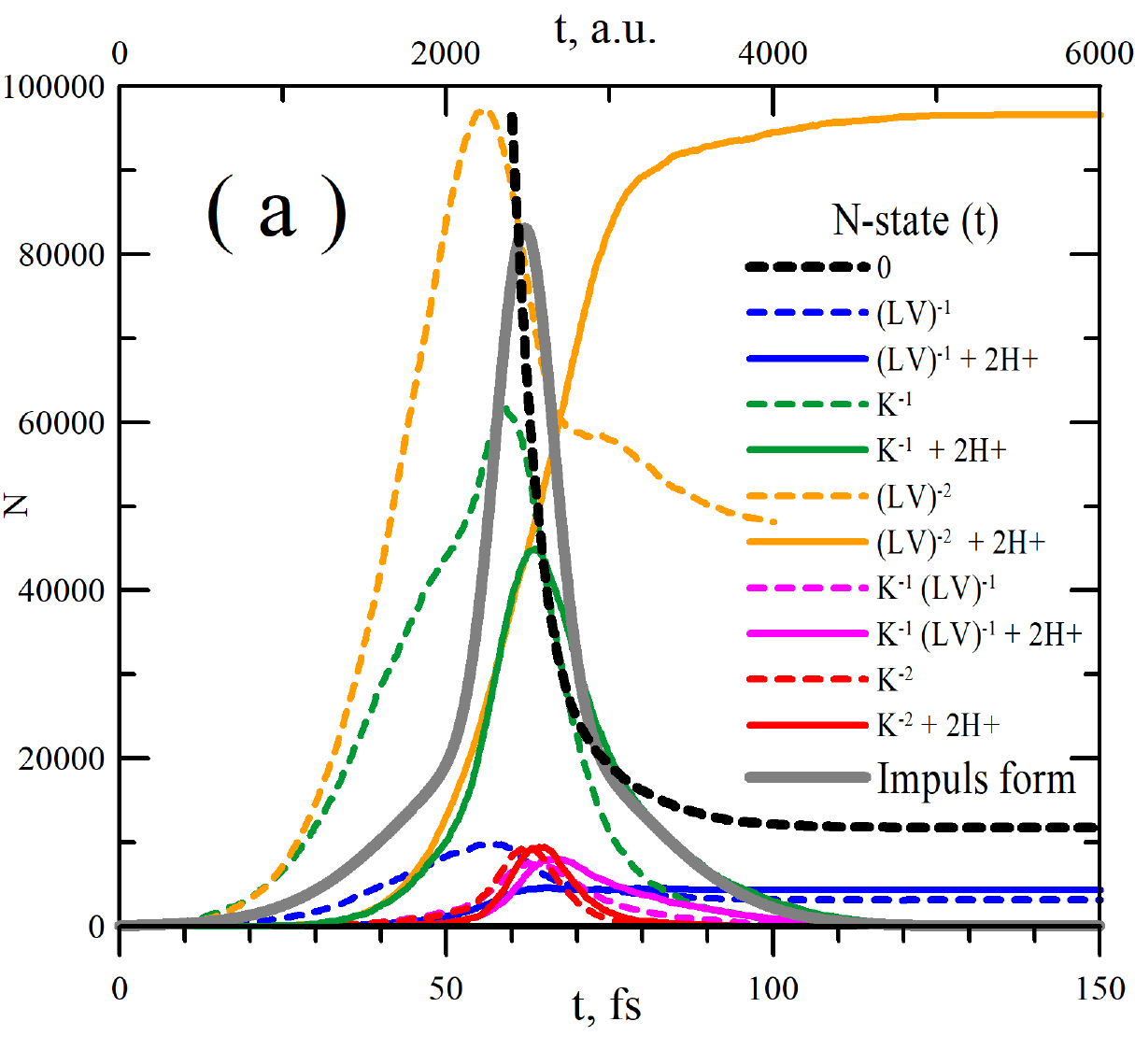}
\includegraphics[width=0.4\textwidth]{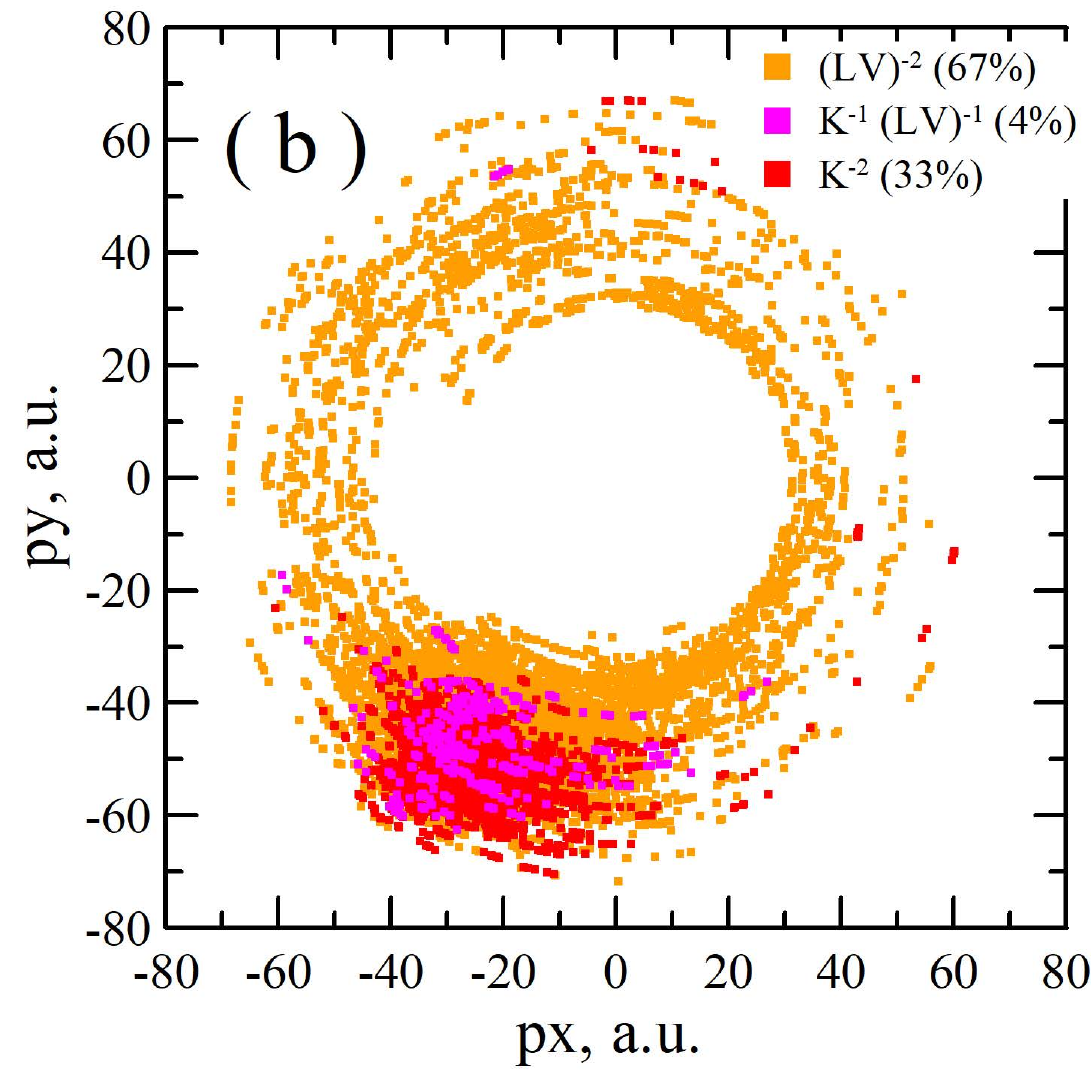}
\caption{Population of the different charge configuration of water fragments as a function of time (a) and the Newton diagram with indication of configuration at the moment the system gets charge $+2$ (b). The pulse parameters as defined by (\ref{eq:pulse2}). Different lines and points mark neutral water (black), cation in the ground state (blue), cation with SCH (green), dication in the ground state (orange), with SCH (magenta) and DCH (red).}
\label{fig:TD_states}
\end{center}
\end{figure}

Ionization of a molecule by synchrotron radiation causes various chains of events, which we call {\it  configuration trajectories}. The ionization of inner $K$-shell dominates over valence-shell ionization, and afterward a competition between Auger ($\tau_{Au}\approx 5$ fs) and subsequent photo\-ionization ($\tau_ph\approx 1/I\sigma\approx2-5$~fs) starts.
In Fig.~\ref{fig:TD_states}a, there are configurations' populations as a function of time. At the end of the pulse, the dominant contribution originates from O$^{2+}$ ions either with ionized ($Z=+4$) or neutral ($Z=+2$) hydrogen atoms. Let us emphasize that overall number of trajectories for this calculation is $\approx$430000, while the sum of populations at the end of the pulse is around $\approx$170000, therefore, more than half of events are further ionized to higher charged states.
A kind of shoulder in SCH population (green dashed line) on the left side results from competition between depletion of the ground state,  on one hand, and increasing pulse intensity, on the other hand.  For the chosen pulse parameters Eq.~(\ref{eq:pulse2}), the trajectories through SCH occur three times more often than through DCH (Fig.~\ref{fig:TD_states}b), which decay very quickly leading to a hump in the population of the dication  ground state (orange dashed line). The ratio between  SCH and DCH indicates that the intensity is not high, nevertheless, it is enough to provide depletion of long-living ground state cation (V$^{-1}$ dashed blue line) and dication (V$^{-2}$).

While the geometry of the H$_2$O$^{+}$ slightly differs from the geometry of a neutral molecule, regardless of whether ionization has occurred from the internal or a valence shell, the further trajectory, on the contrary, critically depends on where the vacancy occurs. There are two trajectories leading to a more stable dication with the following rich dynamics:
 the $K$-shell ionization with Auger decay and  two subsequent ionization events of valence shell. The configuration of  dication (V$^{-2}$) is unfolding to the linear geometry ($\tau\approx 12-20$~fs). On the contrary, creation of DCH instantaneously leads to fragmentation. Following this logic, we plotted the molecular dynamics with an indication of the event type (Figs.~\ref{fig:TD_states}b, \ref{fig:fig_all_DCH}d-f).

 A characteristic and important feature of the measurements reported in \cite{Jahnke} is a noticeable amount of protons detected in the same hemisphere with an oxygen ion. For this to become possible, the proton must detach from a highly unfolded configuration of the molecular ion (the bond angle is about  180$^{\circ}$) and at the same time the ion must be significantly asymmetric. Then the ''distant'' proton H$^+$ dissociates afterward remaining  ion HO$^+$ is polarized in a way to align oxygen in the direction of the first proton.  Asymmetric vibrations are automatically taken into account in calculations with quantum mechanical PES, but can be involved only via initial (boundary)  conditions.

In Fig.~\ref{fig:TD_states}b, the Newton diagram for one of protons is presented with differentiation between configuration trajectories. The long living dication in $(LV)^{-2}$ configuration manifests brighter dynamical effects, most of events along $x$-axis are caused by them. DCH and rare events associated with formation of $K^{-1}(LV)^{-1}$ are primarily localized closer to initial distribution. Note that the inner structure corresponding to the lowest energy protons is not a ring but rather "fish" with tail oriented opposite to oxygen, because first departed proton obtains less energy.

\subsection{Pulse duration effect and the role of DCH events}

\begin{figure*}[tbh]
\begin{center}
\includegraphics[width=0.9\textwidth]{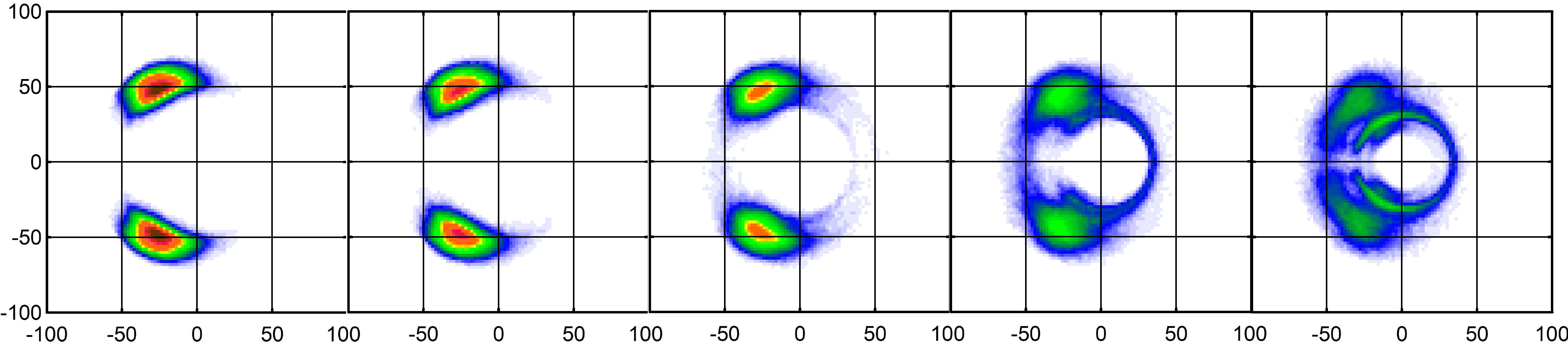}
\caption{The Newton diagram for protons detected in the coincidence with doubly charged oxygen ion O$^{+2}$ for the different pulse durations: FWHM=5, 10, 20, 50 and 100 fs.}
\label{fig:Newton_Time}
\end{center}
\end{figure*}

In Fig.~\ref{fig:Newton_Time}, there are the Newton diagrams for different pulse durations, provided that the total fluence is constant. We fix fluence to $F=3\cdot 10^{11} ph/mkm^2$ in order to keep the probability of ionization approximately constant, therefore, the longer pulse the lower instant intensity. When the pulse is short and instant intensity is high, ionization occurs quickly with essential contribution of DCH events and Newton diagram inherits the geometry of neutral molecule with fluctuation  blur (a); longer duration makes possible emission of a proton to the same hemisphere as an oxygen ion (c); with further increase in duration, proton emission in the opposite direction becomes noticeable; finally, for a sufficiently long pulse, the inner ring corresponding to dissociation with the minimal energy release appears (e).

\begin{figure*}[tbh]
\begin{center}
\includegraphics[width=0.9\textwidth]{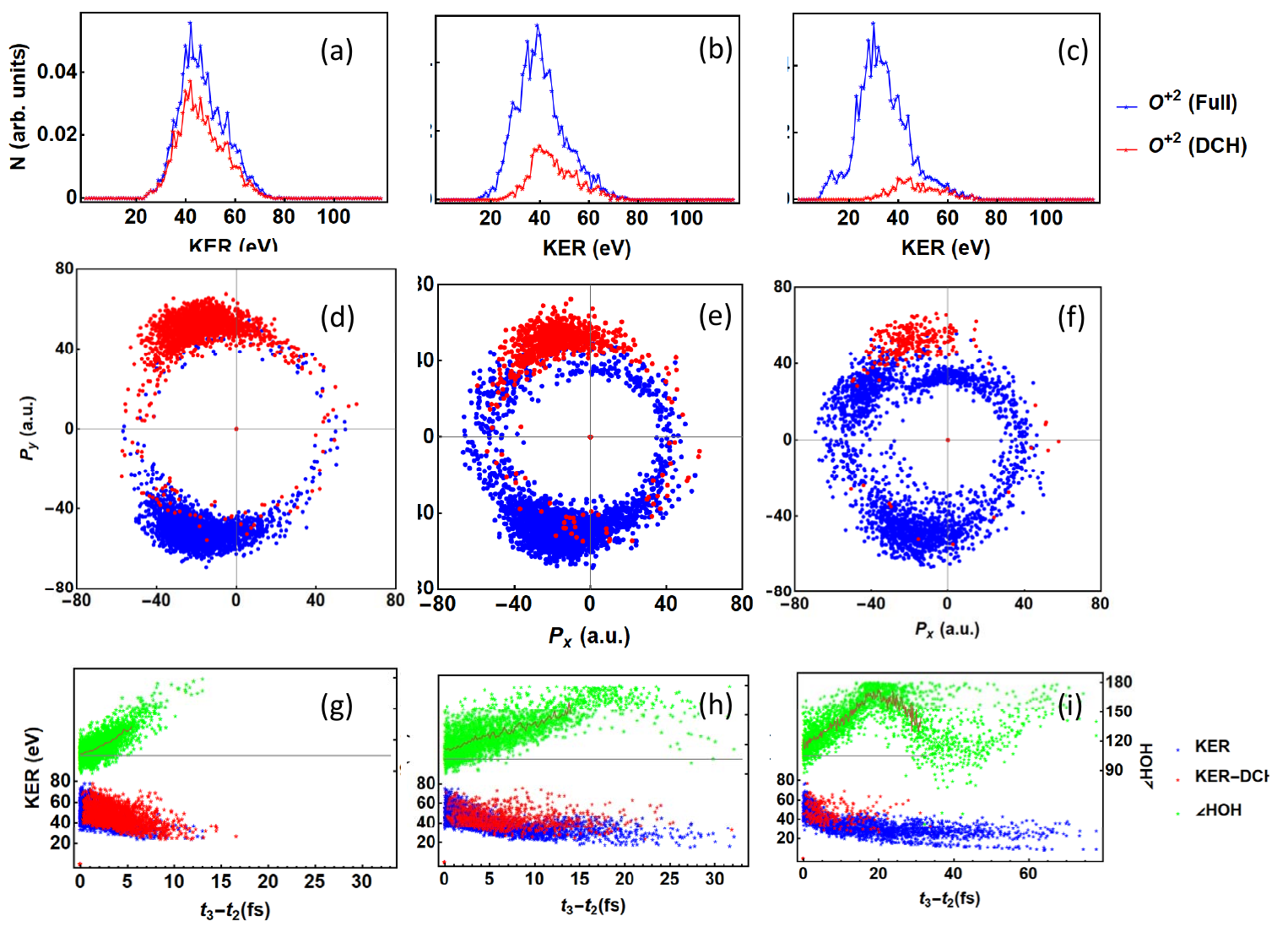}
\caption{Kinetic energy release (a-c), the Newton diagrams (d-f), and time and angle correlations (g-i) for three values of FWHM: 5 fs (a,d,g), 20 fs (b,e,h) and 50 fs (c,f,i). Blue lines and dots correspond to the total number of events (for one proton), the red ones --- to the events proceeding via DCH (for the second proton). }
\label{fig:fig_all_DCH}
\end{center}
\end{figure*}

In Fig. \ref{fig:fig_all_DCH}, the detailed dynamics of water fragmentation is presented with indication of DCH events for different pulse durations (FWHM = 5 fs, 20 fs and 50 fs).  We present KER; the Newton diagrams plotted in a specific way, i.e. for the one proton, we show DCH events only and for the remaining proton, we show all events; and time- and angle-correlations, namely, a time between  a dication gets charge $+3$ and an angle between protons corresponding this event (as charge $+3$ means instant dissociation, it can be understood as {\it angle of fragmentation}). The percentage of DCH events reaches $\approx 70\%$ , 30\%  and 10\% for FWHM=5~fs, 20~fs  and 50~fs, correspondingly.

For the short pulse (a,d,g), DCH events dominate and the KER for all events combined is expected to follow the KER for DCH events only. All-included and DCH-only correlation intervals  practically coincide (Fig.~\ref{fig:fig_all_DCH}g); and the angle of fragmentation hardly reaches $130^{\circ}$.

For the long pulse (c,f,i), events without formation of DCH contribute more in the lower energy region (Fig.~\ref{fig:fig_all_DCH}c), because trajectories $K$-shell ionization---Auger lead to a more stable water dication  which dissociates with minimal energy release. These low energetic events form inner ring appearing in Fig.~\ref{fig:fig_all_DCH}e and especially pronounced in Fig.~\ref{fig:fig_all_DCH}f. Independently on pulse duration, the bond angle reaches $180^\circ$ at time $\approx 20$~fs (Fig.~\ref{fig:fig_all_DCH}h,i), and that corresponds to the time of dication unfolding obtained from PES calculation. For the middle duration, the angle $180^\circ$ corresponds to the end of fragmentation (Fig.~\ref{fig:fig_all_DCH}h), while for longer pulse (Fig.~\ref{fig:fig_all_DCH}i), water dication may complete one full oscillation (around 40 fs) and start the second one.
For the long pulse, the additional structure appears in KER: one can see at least three different energy tails in Fig.~\ref{fig:fig_all_DCH}i corresponding to dissociation.

\widetext
\endwidetext

\subsection{Dissociation dynamics at higher ionic charge}

\begin{figure}[h!]
\begin{center}
\includegraphics[width=0.4\textwidth]{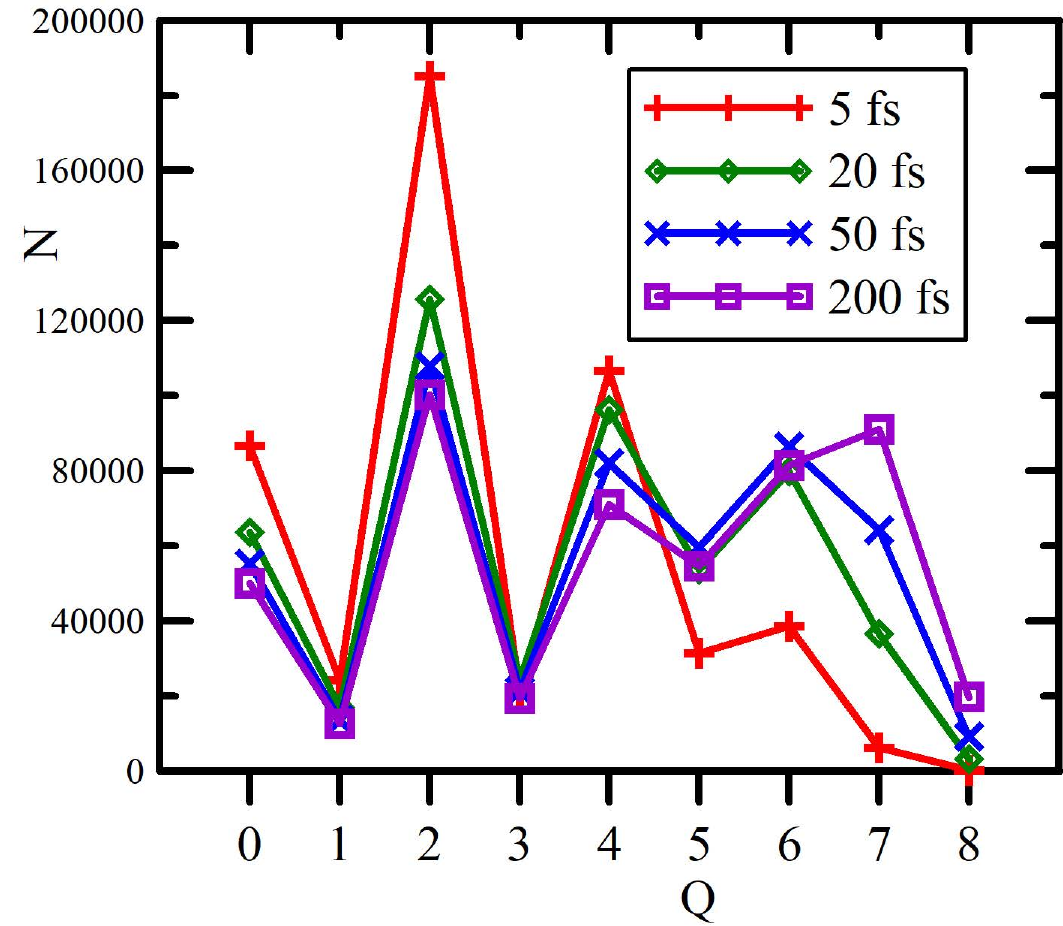}
\caption{Population of the different charge configuration of water ions for different pulse duration (\ref{eq:pulse}).}
\label{fig:Charge}
\end{center}
\end{figure}

The results of our calculations of oxygen ions yields 
are shown in Fig.~\ref{fig:Charge}. The previously noted \cite{Buth} preferential formation of ions with even charges is observed, since the trajectories with ionization of the inner shell and subsequent Auger decay are realized with a higher probability.  The slightly greater in comparison with \cite{Jahnke} formation of ions with high charge states $Z=5 \div 8$, indicates, apparently, a slightly longer effective pulse duration  or lower fluence. Highly charged ions are created more efficiently if the pulse is longer with lower intensity but the same fluence.

\begin{figure*}[tbh]
\begin{center}
\includegraphics[width=0.2\textwidth]{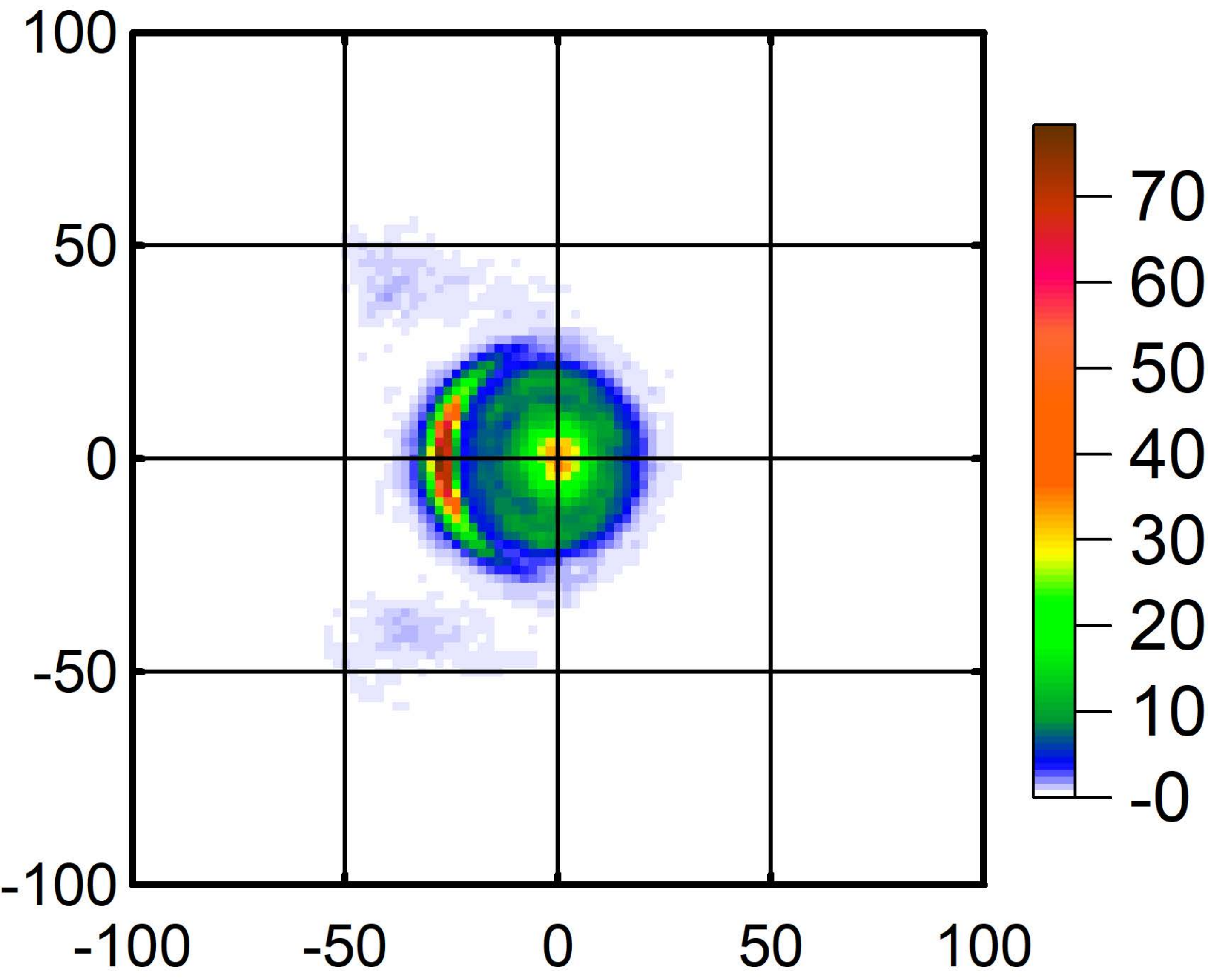}
\includegraphics[width=0.2\textwidth]{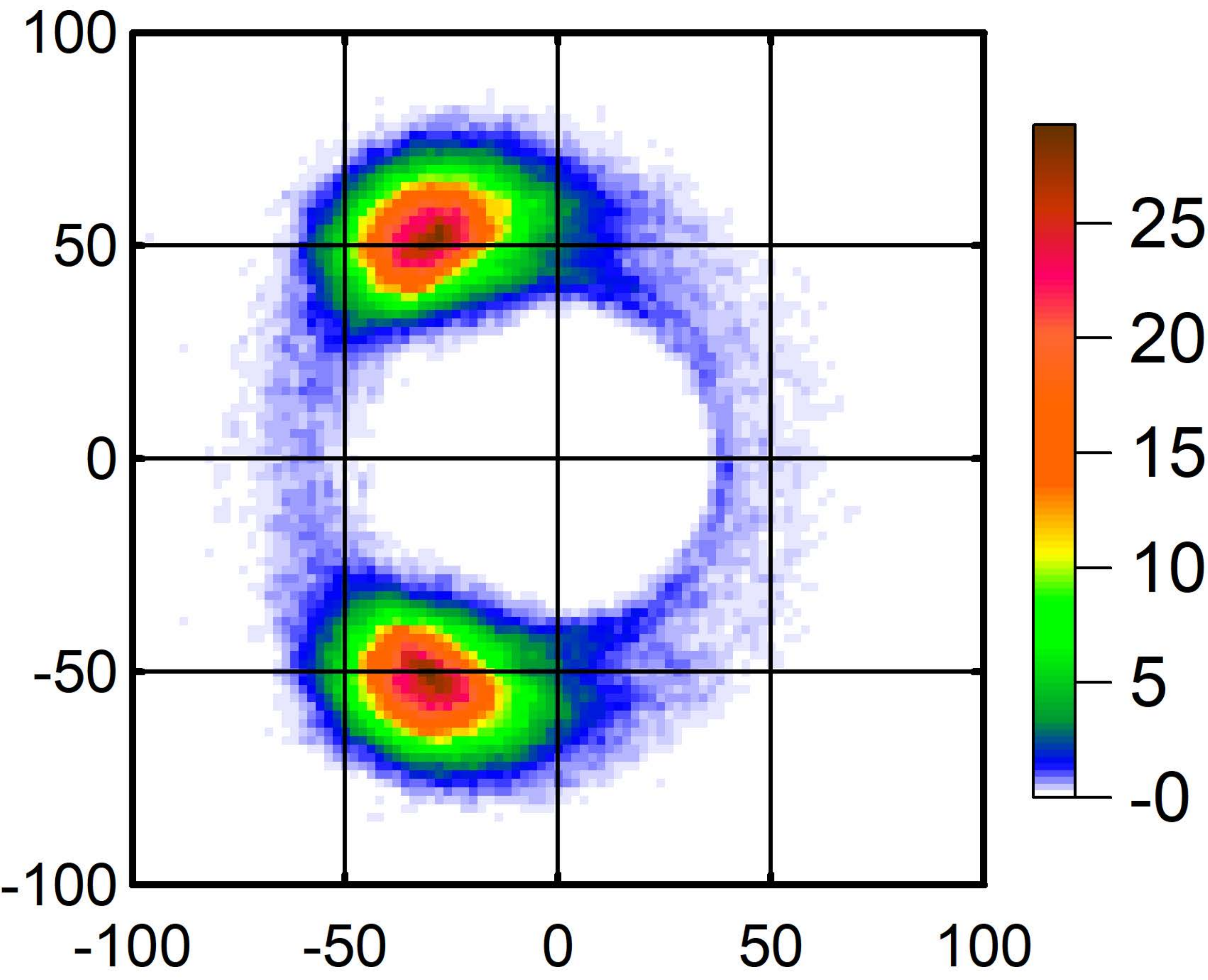}
\includegraphics[width=0.2\textwidth]{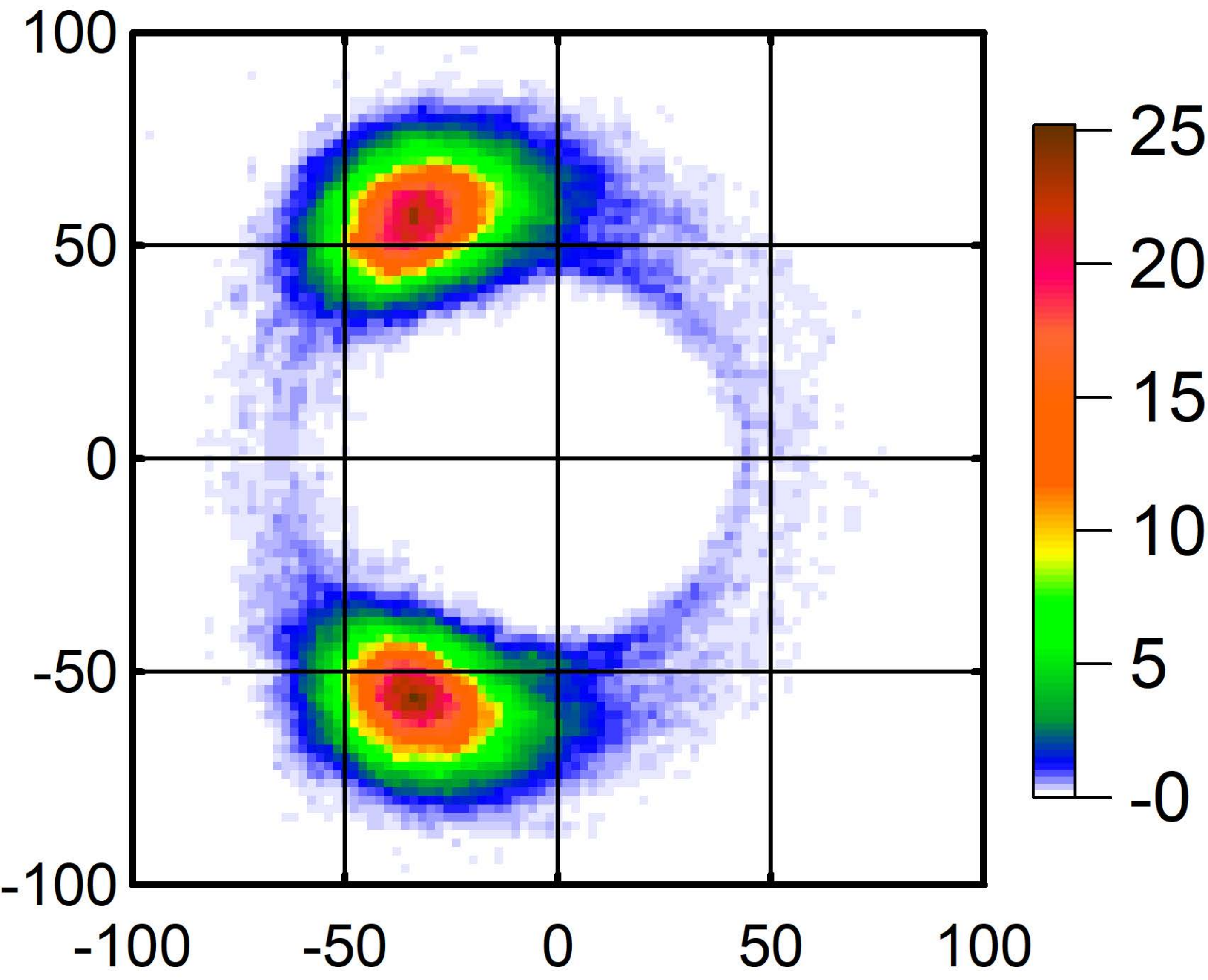}
\includegraphics[width=0.2\textwidth]{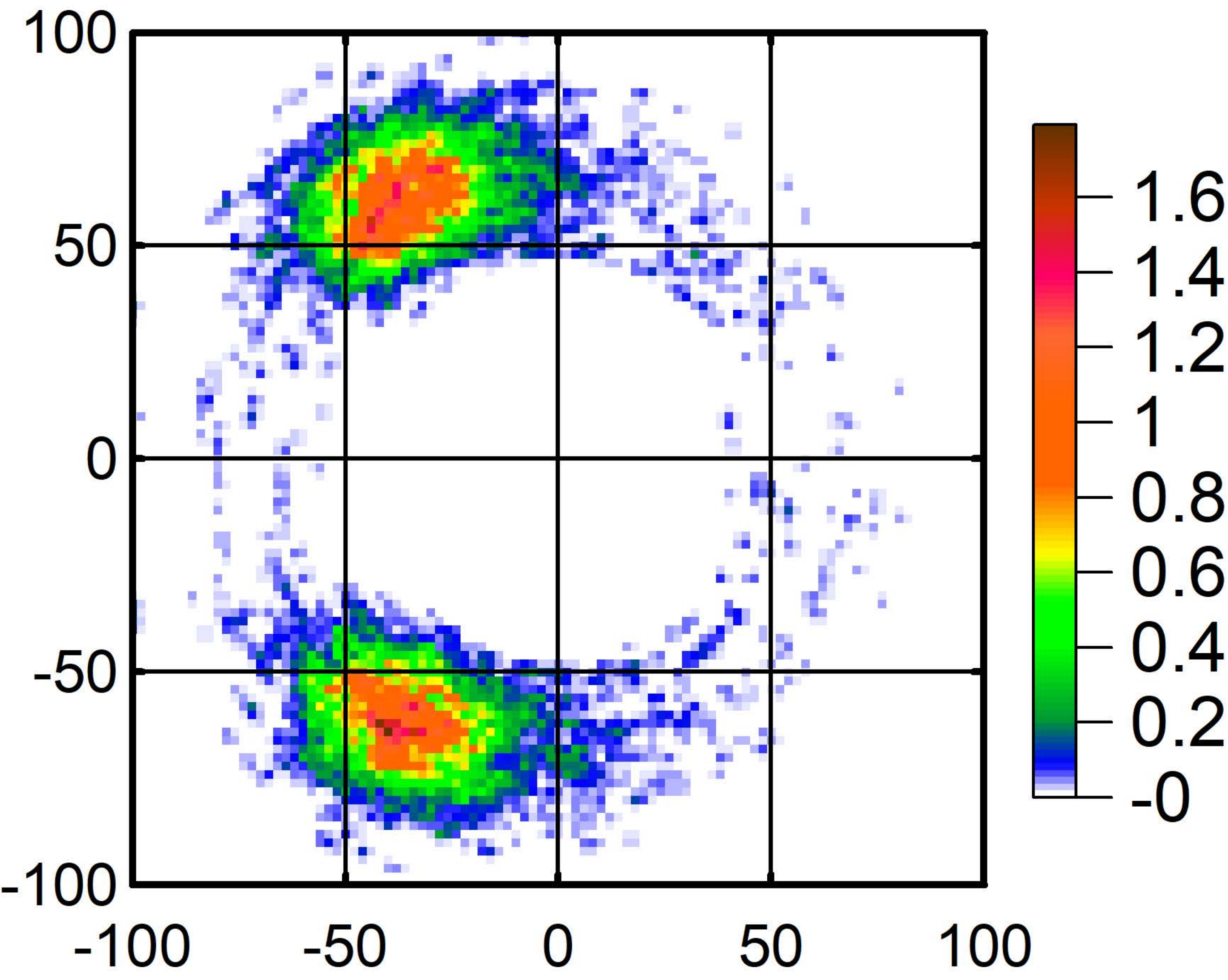}
\caption{The Newton diagram for protons detected in a coincidence with neutral oxygen and its ions.
}
\label{fig:charges}
\end{center}
\end{figure*}

To study the difference between early and late dissociation of a molecule, in Fig.~\ref{fig:charges}, we have provided a comparison of Newton diagrams for protons measured in coincidence with neutral oxygen and its ions O$^{+2}$, O$^{+ 4}$, and O$^{+6}$. The oxygen ions in these charge states prevail at the end of the pulse. Obviously, events with a higher charge correspond to earlier ionization of neutral water and faster decay of the molecular ion. That leads to a decrease in the number of protons registered in the right hemisphere and an increase of the average momentum of protons, which was observed in the experiment and confirmed by the calculations. Events with neutral oxygen ion are the only way to obtain events inside  the dissociative ring. Moreover, only for them, there is a maximum in the direction opposite to $n_x$.

\widetext
\endwidetext

\section{Conclusion}

In the paper  we considered the dynamics of water molecule fragments initiated by interaction with an intense electromagnetic pulse. The charge distribution of oxygen ions  and the acquired final momenta of protons and oxygen ions at the end of the pulse are simulated. The results are presented in the form of Newton diagrams, that is, the momentum of protons measured in the coincidence with the momentum of oxygen. The kinetic energy release acquired due to the Coulomb explosion of the molecule is calculated. All calculations were performed for conditions relevant to the \cite{Jahnke} experiment, and compared with the results presented there. The agreement of the experimental data and calculations shows the applicability of the proposed method.

The calculated and measured Newton diagrams show maxima  and a tail of events corresponding to a proton emitted in the same hemi\-plane as the oxygen ion. While the main maxima are formed due to Coulomb three-particle scattering, the tail is formed by various vibrational modes of a neutral water molecule, its cation and dication. Calculations have shown a significant sensitivity of the acquired kinetic energy to the shape of the electromagnetic pulse.

Varying the pulse duration allowed us to look at the dynamics of molecular bonds in more depth:
DCH events exhibit less diverse dynamics, while the processes going through the valence shell involve significant rearrangement of the molecule.
The role of pulse duration itself is investigated: with an increase of the pulse duration, (a) more ions of odd charges appear; (b) the fragments tend to have lower kinetic energy and a more isotropic distribution.

The presented method now may be considered as well verified and is supposed to be applicate to more complicate molecules.

\section*{ACKNOWLEDGMENTS}
The study was conducted under the state assignment of Lomonosov Moscow State University. The authors highly appreciate Renaud Guillemin and Marc Simon for the experimental data.




\bibliography{references}

\end{document}